\documentclass[11pt]{amsart}

\usepackage{multibib}
\newcites{ltex}{Books and Reviews}

\usepackage{amsmath,amsfonts,amssymb,latexsym}
\usepackage{array,delarray}
\usepackage[dvips]{graphicx}
\usepackage{xspace}
\usepackage{color}
\usepackage[square,sort&compress]{natbib}

\newcommand{\zz}{\mathbf{Z}}
\newcommand{\rr}{\mathbf{R}}

\newcommand{\kk}{k}

\newcommand{\calc}{\mbox{${\mathcal C}$}}

\newcommand{\cali}{\mbox{${\mathcal I}$}}

\newcommand{\sds}{\mbox{\it SDS}}

\newtheorem{theorem}{Theorem}[section]   
\newtheorem{proposition}[theorem]{Proposition}  
\theoremstyle{definition}
\newtheorem{definition}[theorem]{Definition}   
\theoremstyle{remark}
\newtheorem{remark}[theorem]{Remark}        
\newtheorem{example}[theorem]{Example}        
\numberwithin{equation}{section}     

\long\def\Comment#1{
{
 \endIgnore{\relax} 
 }}
\def\endIgnore{\relax}

\begin{document}

\title{
  A mathematical formalism for agent-based modeling
}

\author[R.~Laubenbacher]{Reinhard~Laubenbacher}
\author[A. S. Jarrah]{Abdul~S.~Jarrah}
\author[H.~Mortveit]{Henning Mortveit}
\address{
Virginia Bioinformatics Institute
Virginia Polytechnic Institute and State University } 

\email[Reinhard~Laubenbacher]{reinhard@vbi.vt.edu}
\email[Abdul~S.~Jarrah]{ajarrah@vbi.vt.edu} 
\email[Henning~Mortveit]{henning@vbi.vt.edu}
\author[S.~S.~Ravi]{S.~S.~Ravi}
\address{Department of Computer Science, University at
Albany--State University of New York} \email[S.~S.~Ravi]{ravi@cs.albany.edu}




\begin{abstract}
Many complex systems can be modeled as multiagent systems in which
the constituent entities (agents) interact with each other. The
global dynamics of such a system is determined by the nature of
the local interactions among the agents. Since it is difficult to
formally analyze complex multiagent systems, they are often
studied through computer simulations. While computer simulations
can be very useful, results obtained through simulations do not
formally validate the observed behavior. Thus, there is a need for
a mathematical framework which one can use to represent multiagent
systems and formally establish their properties. This work
contains a brief exposition of some known mathematical frameworks
that can model multiagent systems. The focus is on one such
framework, namely that of finite dynamical systems. Both,
deterministic and stochastic versions of this framework are
discussed. The paper contains a sampling of the mathematical
results from the literature to show how finite dynamical systems
can be used to carry out a rigorous study of the properties of
multiagent systems and it is shown how the framework can also
serve as a universal model for computation.
\end{abstract}

\maketitle

\tableofcontents
\setcounter{tocdepth}{1}

\section*{Glossary}

{\bf Agent-based simulation.}  An agent-based simulation of a complex system is a
computer model that consists of a collection of agents/variables
that can take on a typically finite collection of states.  The state
of an agent at a given point in time is determined through a
collection of rules that describe the agent's interaction with other
agents.  These rules may be deterministic or stochastic.  The
agent's state depends on the agent's previous state and the state of
a collection of other agents with whom it interacts.

\medskip
{\bf Mathematical framework.} A mathematical framework for
agent-based simulation consists of a collection of mathematical
objects that are considered mathematical abstractions of
agent-based simulations.  This collection of objects should be
general enough to capture the key features of most simulations,
yet specific enough to allow the development of a mathematical
theory with meaningful results and algorithms.

\medskip
{\bf Finite dynamical system.} A finite dynamical system is a
time-discrete dynamical system on a finite state set.  That is, it
is a mapping from a cartesian product of finitely many copies of a
finite set to itself.  This finite set is often considered to be a
field.  Dynamics is generated by iteration of the mapping.

\section{Definition of the subject and its importance}

\def\transims{TRANSIMS\xspace}
\def\cimmsim{\emph{CimmSim}\xspace}

Agent-based simulations are generative or computational approaches
used for analyzing ``complex systems.'' What is a ``system?''
Examples of systems include a collection of molecules in a
container, the population in an urban area, and the brokers in a
stock market. The entities or \emph{agents} in these three systems
would be molecules, individuals and stock brokers, respectively.
The agents in such systems interact in the sense that molecules
collide, individuals come into contact with other individuals and
brokers trade shares. Such systems, often called \emph{multiagent
systems}, are not necessarily complex. The label ``complex'' is
typically attached to a system if the number of agents is large,
if the agent interactions are involved, or if there is a large
degree of heterogeneity in agent character or their interactions.

This is of course not an attempt to define a complex system.
Currently there is no generally agreed upon definition of complex
systems. It is not the goal of this article to provide such a
definition -- for our purposes it will be sufficient to think of a
complex system as a collection of agents interacting in some
manner that involves one or more of the complexity components just
mentioned, that is, with a large number of agents, heterogeneity
in agent character and interactions and possibly stochastic
aspects to all these parts.  The global properties of complex
systems, such as their global dynamics, emerge from the totality
of local interactions between individual agents over time.  While
these local interactions are well understood in many cases, little
is known about the emerging global behavior.  Thus, it is
typically difficult to construct global mathematical models such
as systems of ordinary or partial differential equations, whose
properties one could then analyze.  Agent-based simulations are
one way to create computational models of complex systems that
take their place.

An \emph{agent-based simulation}, sometimes also called an
\emph{individual-based} or \emph{interaction-based simulation}
(which we prefer), of a complex system is in essence a computer
program that realizes some (possibly approximate) model of the
system to be studied, incorporating the agents and their rules of
interaction. The simulation might be deterministic (i.e., the
evolution of agent-states is governed by deterministic rules) or
stochastic. The typical way in which such simulations are used is
to initialize the computer program with a particular assignment of
agent states and to run it for some time. The output is a temporal
sequence of states for all agents, which is then used to draw
conclusions about the complex system one is trying to understand.
In other words, the computer program is the model of the complex
system, and by running the program repeatedly, one expects to
obtain an understanding of the characteristics of the complex
system.

There are two main drawbacks to this approach.  First, it is
difficult to validate the model.  Simulations for most systems
involve quite complex software constructs that pose challenges to
code validation. Second, there are essentially no rigorous tools
available for an analysis of model properties and dynamics.  There
is also no widely applicable formalism for the comparison of
models.  For instance, if one agent-based simulation is a
simplification of another, then one would like to be able to
relate their dynamics in a rigorous fashion.  We are currently
lacking a mathematically rich formal framework that models
agent-based simulations. This framework should have at its core a
class of mathematical objects to which one can map agent-based
simulations. The objects should have a sufficiently general
mathematical structure to capture key features of agent-based
simulations and, at the same time, should be rich enough to allow
the derivation of substantial mathematical results.  This chapter
presents one such framework, namely the class of
\emph{time-discrete dynamical systems over finite state sets}.

 The building blocks of these systems consist of a collection of
variables (mapping to agents), a dependency graph that captures
the dependence relations of agents on other agents, a local update
function for each agent that encapsulates the rules by which the
state of each agent evolves over time, and an update discipline
for the variables (e.g. parallel or sequential). We will show that
this class of mathematical objects is appropriate for the
representation of agent-based simulations and, therefore, complex
systems, and is rich enough to pose and answer relevant
mathematical questions. This class is sufficiently rich to be of
mathematical interest in its own right and much work remains to be
done in studying it. We also remark that many other frameworks
such as probabilistic Boolean networks~\cite{Shmulevich:02b} fit
inside the framework described here.


\section{Introduction}

Computer simulations have become an integral part of today's
research and analysis methodologies. The ever-increasing demands
arising from the complexity and sheer size of the phenomena
studied constantly push computational boundaries, challenge
existing computational methodologies, and motivate the development
of new theories to improve our understanding of the potential and
limitations of computer simulation. Interaction-based simulations
are being used to simulate a variety of biological systems such as
ecosystems and the immune system, social systems such as urban
populations and markets, and infrastructure systems such as
communication networks and power grids.

To model or describe a given system, one typically has several choices
in the construction and design of agent-based models and
representations. When agents are chosen to be simple, the simulation
may not capture the behavior of the real system. On the other hand,
the use of highly sophisticated agents can quickly lead to complex
behavior and dynamics. Also, use of sophisticated agents may lead
to a system that \emph{scales poorly}. That is, a linear increase in
the number of agents in the system may require a non-linear (e.g.,
quadratic, cubic, or exponential) increase in the computational
resources needed for the simulation.

Two common methods, namely \emph{discrete event simulation} and
\emph{time-stepped simulation}, are often used to implement
agent-based models~\cite{Jefferson:85,Bagrodia:98,Nance:93}.  In
the discrete event simulation method, each event that occurs in
the system is assigned a time of occurrence.  The collection of
events is kept in increasing order of their occurrence times.
(Note that an event occurring at a certain time may give rise to
events which occur later.)  When all the events that occur at a
particular time instant have been carried out, the simulation
clock is advanced to the next time instant in the order.  Thus,
the differences between successive values of the simulation clock
may not be uniform.  Discrete event simulation is typically used
in contexts such as queuing systems \cite{Misra:86}.  In the
time-stepped method of simulation, the simulation clock is always
advanced by the same amount.  For each value of the simulation
clock, the states of the system components are computed using
equations that model the system.  This method of simulation is
commonly used for studying, e.g., fluid flows or chemical
reactions. The choice of model (discrete event versus
time-stepped) is typically guided by an analysis of the computational
speed they can offer, which in turn depends on the nature of the
system being modeled, see, e.g.,~\cite{Guo:2000}.

Tool-kits for general purpose agent-based simulations include
Swarm~\cite{Langton:96,Ebeling:01} and Repast~\cite{Repast:06}.
Such tool-kits allow one to specify more complex agents and
interactions than would be possible using, e.g., ordinary
differential equations models.  In general, it is difficult to
develop a software package that is capable of supporting the
simulation of a wide variety of physical, biological, and social
systems.

Standard or classical approaches to modeling are often based on
continuous techniques and frameworks such as ordinary differential
equations (ODEs) and partial differential equations (PDEs). For
example, there are PDE based models for studying traffic
flow~\cite{Whitham:74,Gupta:05,Keyfitz:04}. These can accurately model
the emergence of traffic jams for simple road/intersection
configurations through, for example, the formation of shocks.
However, these models fail to scale to the size and the specifications
required to accurately represent large urban areas.  Even if they
hypothetically were to scale to the required size, the answers they
provide (e.g. car density on a road as a function of position and
time) cannot answer questions pertaining to specific travelers or
cars. Questions of this type can be naturally described and answered
through agent-based models. An example of such a system is \transims
(see Section~\ref{sec:transims}), where an agent-based simulation
scheme is implemented through a cellular automaton model.  Another
well-known example of the change in modeling paradigms from continuous
to discrete is given by lattice gas automata~\cite{Hasslacher:86} in
the context of fluid dynamics.

Stochastic elements are inherent in many systems, and this usually
is reflected in the resulting models used to describe them. A
stochastic framework is a natural approach in the modeling of, for
example, noise over a channel in a simulation of telecommunication
networks~\cite{Barrett:02z}. In an economic market or a game
theoretic setting with competing players, a player may sometimes
decide to provide incorrect information. The state of such a player
may therefore be viewed and modeled by a random variable. A player
may make certain probabilistic assumptions about other players'
environment.  In biological systems, certain features and properties
may only be known up to the level of probability distributions. It
is only natural to incorporate this stochasticity into models of
such systems.

Since applications of stochastic discrete models are common, it is
desirable to obtain a better understanding of these simulations
both from an application point of view (reliability, validation)
and from a mathematical point of view. However, an important
disadvantage of agent-based models is that there are few
mathematical tools available at this time for the analysis of
their dynamics.


\subsection{Examples of Agent-Based Simulations}

In order to provide the reader with some concrete examples that
can also be used later on to illustrate theoretical concepts we
describe here three examples of agent-based descriptions of
complex systems, ranging from traffic networks to the immune
system and voting schemes.


\subsubsection{\transims{} (\emph{TR}ansportation
\emph{AN}alysis \emph{SIM}ulation \emph{S}ystem)}
\label{sec:transims}

\transims is a large-scale computer simulation of traffic on a
road network~\cite{Nagel:97a,Nagel:96a,Nagel:06}. The simulation
works at the resolution level of individual travelers, and has
been used to study large US metropolitan areas such as
Portland,OR, Washington D.C. and Dallas/Fort Worth. A
\transims-based analysis of an urban area requires $(i)$ a
population, $(ii)$ a location-based activity plan for each
individual for the duration of the analysis period and $(iii)$ a
network representation of all transportation pathways of the given
area. 
The data required for $(i)$ and $(ii)$ are generated based on,
e.g., extensive surveys and other information sources. The network
representation is typically very close to a complete description
of the real transportation network of the given urban area.

\transims{} consists of two main modules: the \emph{router} and
the cellular automaton based \emph{micro-simulator}. The router
maps each activity plan for each individual (obtained typically
from activity surveys) into a travel route. The micro-simulator
executes the travel routes and sends each individual through the
transportation network so that its activity plan is carried out.
This is done in such a way that all constraints imposed on
individuals from traffic driving rules, road signals, fellow
travelers, and public transportation schedules are respected. The
time scale is typically $1$ second.

The micro-simulator is the part of \transims responsible for the
detailed traffic dynamics. Its implementation is based on
\emph{cellular automata} which are described in more detail in
Section~\ref{sec:CA}. Here, for simplicity, we focus on the
situation where each individual travels by car. The \emph{road
network representation} is in terms of \emph{links} (e.g.~road
segments) and \emph{nodes} (e.g.~intersections). 
The network description is turned into a cell-network description
by discretizing each lane of every link into cells. A cell
corresponds to a $7.5$ meter lane segment, and can have up to four
neighbor cells (front, back, left and right).

The \emph{vehicle dynamics} is specified as follows. Vehicles
travel with discrete velocities $0$, $1$, $2$, $3$, $4$ or $5$
which are constant between time steps. Each update time step
brings the simulation one time unit forward. If the time unit is
one second then the maximum speed of $v_{\max} = 5$ cells per time
unit corresponds to an actual speed of $5\times7.5\,\mathrm{m/s} =
37.5\,\mathrm{m/s}$ which is $135\,\mathrm{kmh}$ or approximately
$83.9\,\mathrm{mph}$.

Ignoring intersection dynamics, the micro-simulator executes three
functions for each vehicle in every update: $(a)$ lane-changing,
$(b)$ acceleration and $(c)$ movement.  These functions can be
implemented through four cellular automata, one each for lane
change decision and execution, one for acceleration and one for
movement. For instance, the acceleration automaton works as
follows. A vehicle in \transims can increase its speed by at most
$1$ cell per second, but if the road ahead is blocked, the vehicle
can come to a complete stop in the same time.  The function that
is applied to each cell that has a car in it uses the gap ahead
and the maximal speed to determine if the car will increase or
decrease its velocity. Additionally, a car may have its velocity
decreased one unit as determined by a certain deceleration
probability. The random deceleration is an important element of
producing realistic traffic flow. A major advantage of this
representation is that it leads to very light-weight agents, a
feature that is critical for achieving efficient scaling.


\subsubsection{CImmSim}

Next we discuss an interaction-based simulation that models certain
aspects of the human immune system. Comprised of a large number of
interacting cells whose motion is constrained by the body's anatomy,
the immune system lends itself very well to agent-based simulation. In
particular, these models can take into account three-dimensional
anatomical variations as well as small-scale variability in cell
distributions. For instance, while the number of T-cells in the human
body is astronomical, the number of antigen-specific T-cells, for a
specific antigen, can be quite small, thereby creating many spatial
inhomogeneities.  Also, little is known about the global structure of
the system to be modeled.

The first discrete model to incorporate a useful level of
complexity was \emph{ImmSim}~\citep{CS1,CS2}, developed by Seiden
and Celada as a stochastic cellular automaton.  The system
includes B-cells, T-cells, antigen presenting cells (APCs),
antibodies, antigens, and antibody-antigen complexes.  Receptors
on cells are represented by bit strings, and antibodies use bit
strings to represent their epitopes and peptides. Specificity and
affinity are defined by using bit string similarity. The bit
string approach was initially introduced in \citep{F}.  The model
is implemented on a regular two-dimensional grid, which can be
thought of as a slice of a lymph node, for instance. It has been
used to study various phenomena, including the optimal number of
human leukocyte antigens in human beings~\citep{CS1}, the
autoimmunity and T-lymphocyte selection in the thymus~\citep{M},
antibody selection and hyper-mutation~\citep{CS3}, and the
dependence of the selection and maturation of the immune response
on the antigen-to-receptor's affinity~\citep{BC1}. The
computational limitations of the Seiden-Celada model have been
overcome by a modified model, \cimmsim~\cite{cimmsim}, implemented
on a parallel architecture. Its complexity is several orders of
magnitude larger than that of its predecessor. It has been used to
model hypersensitivity to chemotherapy~\citep{CA} and the
selection of escape mutants from immune recognition during HIV
infection~\citep{BC2}. In \cite{ebv} the \cimmsim framework was
applied to the study of mechanisms that contribute to the
persistence of infection with the Epstein-Barr virus.


\subsubsection{A Voting Game}
\label{sec:voting}

The following example describes a hypothetical voting scheme.  The
voting system is constructed from a collection of voters. For
simplicity, it is assumed that only two candidates, represented by
$0$ and $1$, contest in the election. There are $N$ voters
represented by the set $\{v_1,v_2,\ldots,v_N\}$. Each voter has a
candidate preference or a \emph{state}. We denote the state of
voter $v_i$ by $x_i$. Moreover, each voter knows the preferences
or states of some of his or her friends (fellow voters). This
friendship relation is captured by the \emph{dependency graph}
which we describe later in Section~\ref{sec:fds-def}.

Starting from an initial configuration of preferences, the voters cast
their votes in some order. The candidate that receives the most votes
is the winner.  A number of rules can be formulated to decide how each
voter chooses a candidate.  We will provide examples of such rules
later, and as will be seen, the outcome of the election is governed by
the order in which the votes are cast as well as the structure of the
dependency graph.



\section{Existing mathematical frameworks}

The field of agent-based simulation currently places heavy emphasis on
implementation and computation rather than on the derivation of formal
results. Computation is no doubt a very useful way to discover
potentially interesting behavior and phenomena. However, unless the
simulation has been set up very carefully, its outcome does not
formally validate or guarantee the observed phenomenon. It could
simply be caused by an artifact of the system model, an implementation
error, or some other uncertainty.

A first step in a theory of agent-based simulation is the introduction
of a formal framework that on the one hand is precise and
computationally powerful, and, on the other hand, is natural in the
sense that it can be used to effectively describe large classes of
both deterministic and stochastic systems. Apart from providing a
common basis and a language for describing the model using a sound
formalism, such a framework has many advantages. At a first level, it
helps to clarify the key structure in a system.  Domain specific
knowledge is crucial to deriving good models of complex systems, but
domain specificity is often confounded by domain conventions and
terminology that eventually obfuscate the real structure.

A formal, context independent framework also makes it easier to take
advantage of existing general theory and algorithms. Having a model
formulated in such a framework also makes it easier to establish
results. Additionally, expressing the model using a general framework
is more likely to produce results that are widely applicable. This
type of framework also supports implementation and
validation. Modeling languages like UML~\cite{Booch:05} serve a
similar purpose, but tend to focus solely on software implementation
and validation issues, and very little on mathematical or
computational analysis.


\subsection{Cellular Automata}\label{sec:CA}

In this section we discuss several existing frameworks for
describing agent-based simulations. Cellular automata (CA) were
introduced by Ulam and von~Neumann~\cite{Neumann:66a} as
biologically motivated models of computation. Early research
addressed questions about the computational power of these
devices. Since then their properties have been studied in the
context of dynamical systems~\cite{Hedlund:69a}, language
theory~\cite{Lindgren:98}, and ergodic theory~\cite{Lind:84} to
mention just a few areas. Cellular automata were popularized by
Conway~\cite{Gardner:70} (Game of Life) and by
Wolfram~\cite{Martin:84,Wolfram:83,Wolfram:02}. Cellular automata
(both deterministic and stochastic) have been used as models for
phenomena ranging from lattice gases~\cite{Hasslacher:86} and
flows in porous media~\cite{Rothman:88} to traffic
analysis~\cite{Nagel:95a,Nagel:92a,Fuks:04}.

A cellular automaton is typically defined over a regular grid. An
example is a two-dimensional grid such as $\zz^2$. Each grid point
$(i,j)$ is referred to as a \emph{site} or \emph{node}.  Each site
has a state $x_{i,j}(t)$ which is often taken to be binary. Here $t$
denotes the time step. Furthermore, there is a notion of a
\emph{neighborhood} for each site. The neighborhood $N$ of a site is
the collection of sites that can influence the future state of the
given site. Based on its current state $x_{i,j}(t)$ and the current
states of the sites in its neighborhood $N$, a function $f_{i,j}$ is
used to compute the next state $x_{i,j}(t+1)$ of the site $(i,j)$.
Specifically, we have
\begin{equation}
\label{eq:ca}
 x_{i,j}(t+1) = f_{i,j}(\bar{x}_{i,j}(t))\;,
\end{equation}
where $\bar{x}_{i,j}(t)$ denotes the tuple consisting of all the
states $x_{i', j'}(t)$ with $(i',j') \in N$.  The tuple consisting
of the states of all the sites is the CA \emph{configuration} and is
denoted $x(t) = (x_{i,j}(t))_{i,j}$.  Equation~\eqref{eq:ca} is used
to map the configuration $x(t)$ to $x(t+1)$. The cellular automaton
map or dynamical system, is the map $\Phi$ that sends $x(t)$ to
$x(t+1)$.

A central part of CA research is to understand how configurations
evolve under iteration of the map $\Phi$ and what types of
dynamical behavior can be generated. A general introduction to CA
can be found in~\cite{Ilachinski:01}.

\subsection{Hopfield Networks}
\label{sec:hopfield_net}

Hopfield networks were proposed as a simple model of associative
memories \cite{Hop:82}.  A discrete Hopfield neural network consists
of an undirected graph $Y(V,E)$.  At any time $t$, each node $v_i
\in V$ has a state $x_i(t) \in \{+1, -1\}$.  Further, each node $v_i
\in V$ has an associated \emph{threshold} $\tau_i \in \rr$.  Each
edge $\{v_i, v_j\} \in E$ has an associated weight $w_{i,j} \in
\rr$. For each node $v_i$, the neighborhood $N_i$ of $v_i$ includes
$v_i$ and the set of nodes that are adjacent to $v_i$ in $Y$.
Formally,
\[
N_i = \{v_i\} \cup \{v_j \in V \: : \: \{v_i, v_j\} \in E\}.
\]
States of nodes are updated as follows.  At time $t$, node $v_i$
computes the function $f_i$ defined by
\[
f_i(t) = \mathrm{sgn}\left( - \tau_i + \sum_{v_j \in
N_i}{w_{i,j}\,x_j(t)}
                              \right),
\]
where $\mathrm{sgn}$ is the map from $\rr$ to $\{+1, -1\}$, defined
by 
\[
\mathrm{sgn}(x) = \left\{%
\begin{array}{ll}
    1, & \hbox{ if \,} x \geq 0 \mbox{ and }\\
    -1 &  \mbox{ otherwise}.\\
\end{array}%
\right.
\]
Now, the state of $v_i$ at time $t+1$ is
\[
x_i(t+1) = f_i(t).
\]


Many references on Hopfield networks (see for example
\cite{Hop:82,RN03}) assume that the underlying undirected graph is
complete; that is, there is an edge between every pair of nodes. In
the definition presented above, the graph need not be complete.
However, this does not cause any difficulties since the missing edges
can be assigned weight 0. As a consequence, such edges will not play
any role in determining the dynamics of the system. Both synchronous
and asynchronous update models of Hopfield neural networks have been
considered in the literature. For theoretical results concerning
Hopfield networks see \cite{Orp:94,Orp:96} and the references cited
therein. Reference \cite{RN03} presents a number of applications of
neural networks. In \cite{hopfield_socialogy}, a Hopfield model is
used to study polarization in dynamic networks.


\subsection{Communicating Finite State Machines}

The model of communicating finite state machines (CFSM) was proposed
to analyze protocols used in computer networks.  In some of the
literature, this model is also referred to as ``concurrent transition
systems'' \cite{GC86}.

In the CFSM model, each agent is a process executing on some node
of a distributed computing system.  Although there are minor
differences among the various CFSM models proposed in the
literature~\cite{BZ83,GC86}, the basic set-up models each process
as a finite state machine (FSM).  Thus, each agent is in a certain
state at any time instant $t$.  For each pair of agents, there is
a bidirectional channel through which they can communicate.  The
state of an agent at time $t+1$ is a function of the current state
and the input (if any) on one or more of the channels.  When an
agent (FSM) undergoes a transition from one state to another, it
may also choose to send a message to another agent or receive a
message from an agent. In general, such systems can be synchronous
or asynchronous.  As can be seen, CFSMs are a natural formalism
for studying protocols used in computer networks.  The CFSM model
has been used extensively to prove properties (e.g. deadlock
freedom, fairness) of a number of protocols used in practice (see
\cite{BZ83,GC86} and the references cited therein).

Other frameworks include interacting particle
systems~\cite{Liggett}, and Petri nets~\cite{PetriNets}. There is
a vast literature on both, but space limitations prevent a
discussion here.


\section{Finite Dynamical Systems}

Another, quite general, modeling framework that has been proposed
is that of \emph{finite dynamical systems}, both synchronous and
asynchronous. Here the proposed mathematical object representing
an agent-based simulation is a time-discrete dynamical system on a
finite state set. The description of the systems is modeled after
the key components of an agent-based simulation, namely agents,
the dependency graph, local update functions, and an update order.
This makes a mapping to agent-based simulations natural.  In the
remainder of this chapter we will show that finite dynamical
systems satisfy our criteria for a good mathematical framework in
that they are general enough to serve as a broad computing tool
and mathematically rich enough to allow the derivation of formal
results.

\subsection{Definitions, Background, and Examples}
\label{sec:fds-def}

\def\vname{x}

Let $x_1, \ldots, x_n$ be a collection of variables, which take
values in a finite set $X$. (As will be seen, the variables
represent the entities in the system being modeled and the
elements of $X$ represent their states.) Each variable $x_i$ has
associated to it a ``local update function'' $f_i:
X^n\longrightarrow X$, where ``local" refers to the fact that
$f_i$ takes inputs from the variables in the ``neighborhood" of
$x_i$, in a sense to be made precise below. By abuse of notation
we also let $f_i$ denote the function $X^n\longrightarrow X^n$
which changes the $i$-th coordinate and leaves the other
coordinates unchanged.  This allows for the sequential composition
of the local update functions.  These functions assemble to a
dynamical system
\[
\Phi =(f_1,\dots,f_n): X^n \longrightarrow X^n,
\]
with the dynamics generated by iteration of $\Phi$. As an example,
if $X = \{0, 1\}$ with the standard Boolean operators AND and OR,
then $\Phi$ is a Boolean network.

The assembly of $\Phi$ from the local functions $f_i$ can be done in
one of several ways. One can update each of the variables
simultaneously, that is,
$$
\Phi(\vname_1, \ldots ,\vname_n) = (f_1(\vname_1, \ldots ,\vname_n),\ldots ,f_n(\vname_1, \ldots , \vname_n)) \;.
$$
In this case one obtains a \emph{parallel dynamical system}.

Alternatively, one can choose to update the states of the
variables according to some fixed update order, for example, a
permutation or a word on the set $\{1,\dots, n\}$. That is, let
$\pi=(\pi_1,\dots,\pi_t)$ be a word using the alphabet $\{1,\dots,
n\}$. The function
\begin{equation}\label{seq-fun}
 \Phi_\pi= f_{\pi_t}\circ f_{\pi_{n-1}} \circ\cdots \circ f_{\pi_1} \;,
\end{equation}
is called a \emph{sequential dynamical system (SDS)} and, as
before, the dynamics of $\Phi_\pi$ is generated by iteration. The
case when $\pi$ is a permutation on $\{1,\dots, n\}$ has been
studied extensively \cite{Barrett:99a, Barrett:00a, Barrett:01a,
BMR3}. It is clear that using a different permutation $\gamma$ may
result in a different dynamical system $\Phi_\gamma$.

\begin{remark}\label{seq-para}
Notice that for a fixed $\pi$, the function $\Phi_\pi$ is a
parallel dynamical system: once the update order $\pi$ is chosen
and the local update functions are composed according to $\pi$,
that is, the function $\Phi_\pi$ has been computed, then
$\Phi_\pi(\vname_1,\dots,\vname_n) = g(\vname_1,\dots,\vname_n)$ where $g$ is a
parallel update dynamical system. However, the maps $g_i$ are not
local functions.
\end{remark}

The dynamics of $\Phi$ is usually represented as a directed graph on
the vertex set $X^n$, called the \emph{phase space} of $\Phi$. There
is a directed edge from $\mathbf v\in X^n$ to $\mathbf w\in X^n$ if
and only if $\Phi(\mathbf v) = \mathbf w$.  A second graph that is
usually associated with a finite dynamical system is its
\emph{dependency graph} $Y(V,E)$. In general, this is a directed
graph, and its vertex set is $V=\{1,\dots, n\}$. There is a directed
edge from $i$ to $j$ if and only if $x_i$ appears in the function
$f_j$. In many situations, the interaction relationship between
pairs of variables is symmetric; that is, variable $x_i$ appears in
$f_j$ if and only if $x_j$ appears in $f_i$. In such cases, the
dependency graph can be thought of as an undirected graph. We recall
that the dependency graphs mentioned in the context of the voting
game (Section~\ref{sec:voting}) and Hopfield networks
(Section~\ref{sec:hopfield_net}) are undirected graphs. The
dependency graph plays an important role in the study of finite
dynamical systems and is sometimes listed explicitly as part of the
specification of $\Phi$.

\medskip
\begin{example}\label{run-ex}
Let $X=\{0,1\}$. Suppose we have four variables and the local
Boolean update functions are
\begin{eqnarray*}
  f_1 &=&  x_1+x_2+x_3+x_4,\\
  f_2 &=&  x_1 + x_2,\\
  f_3 &=&  x_1 + x_3,\\
  f_4 &=&  x_1 + x_4,
\end{eqnarray*}
where ``+" represents XOR, the exclusive OR function.  The dynamics
of the function $\Phi=(f_1,\dots,f_4) :X^4 \longrightarrow X^4$ is
the directed graph on the left in Figure \ref{run-ex-para} while the
dependency graph is on the right.
\begin{figure}[ht]
\centerline{ \raise20pt\hbox{ \framebox{
\includegraphics[width=0.4\textwidth]{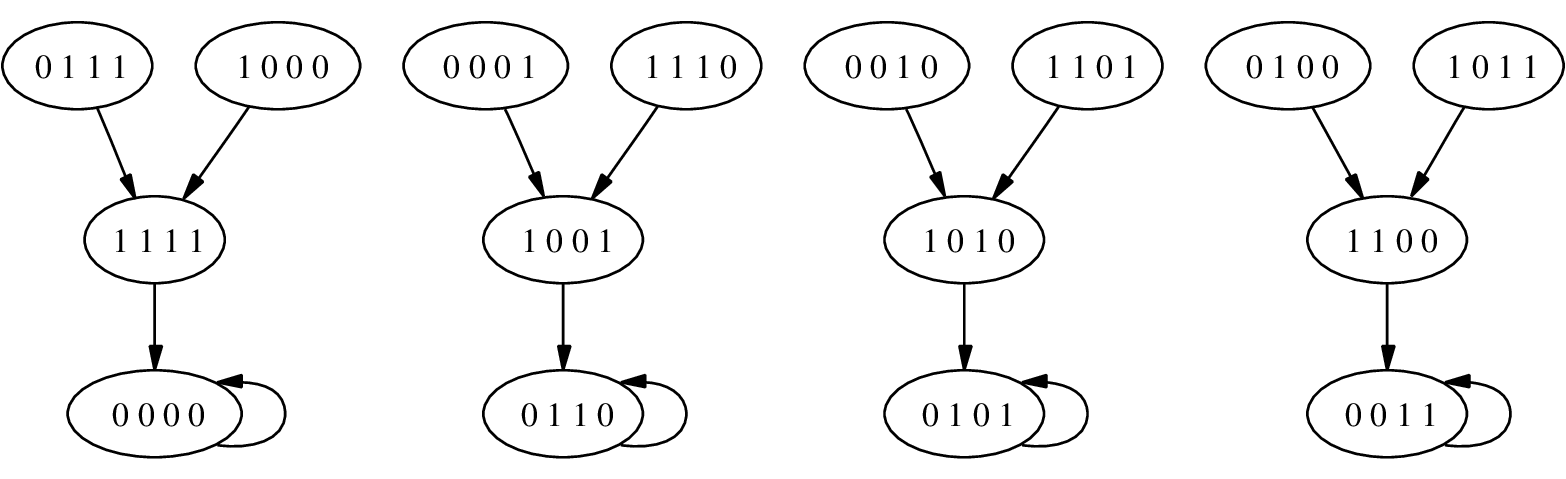}}}
\quad \quad \raise20pt\hbox{ \framebox{
\includegraphics[width=0.25\textwidth]{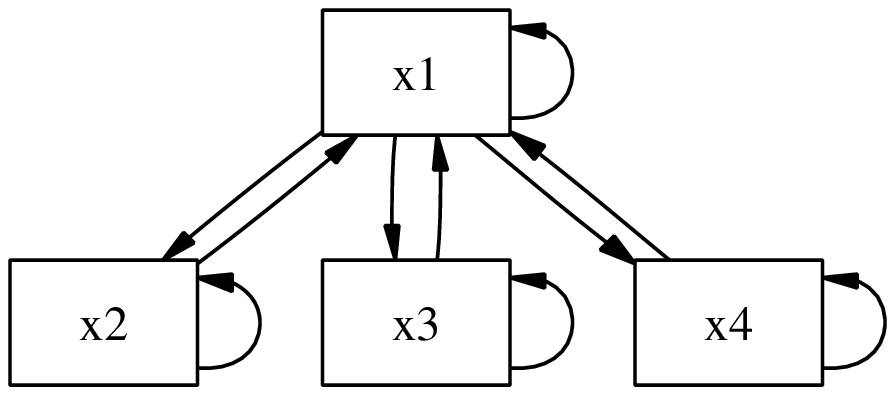}}}
}
\caption{The phase space of the parallel system $\Phi$ is on the left and
dependency graph of the Boolean function from Example \ref{run-ex} is
on the right.}
\label{run-ex-para}
\end{figure}
\end{example}

\begin{example}\label{run-ex-sequential}
Consider the local functions in the Example \ref{run-ex} above and
let $\pi = (2,1,3,4)$. Then
\[
\Phi_\pi = f_4 \circ f_3 \circ f_1 \circ f_2 : X^4 \longrightarrow
X^4.
\]
The phase space of $\Phi_\pi$ is the directed graph on the left in
Figure \ref{run-ex-seq}, while the phase space of $\Phi_\gamma$,
where $\gamma = id$ is on the right in Figure \ref{run-ex-seq}.

\begin{figure}[ht]
\centerline{ \raise20pt\hbox{
\framebox{
\includegraphics[width=0.4\textwidth]{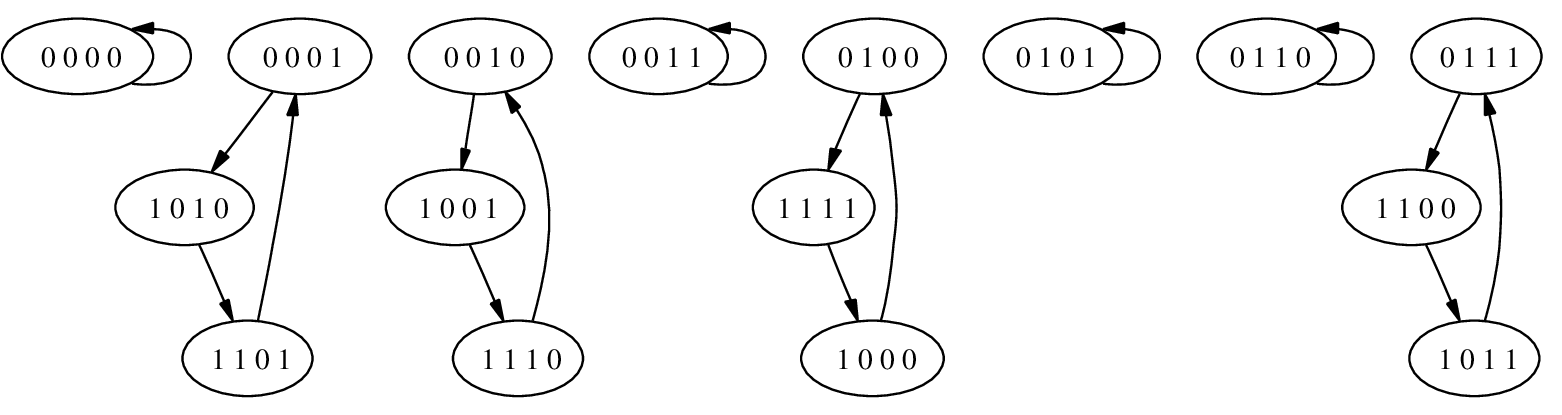}}}
\quad \raise20pt\hbox{
\framebox{
\includegraphics[width=0.4\textwidth]{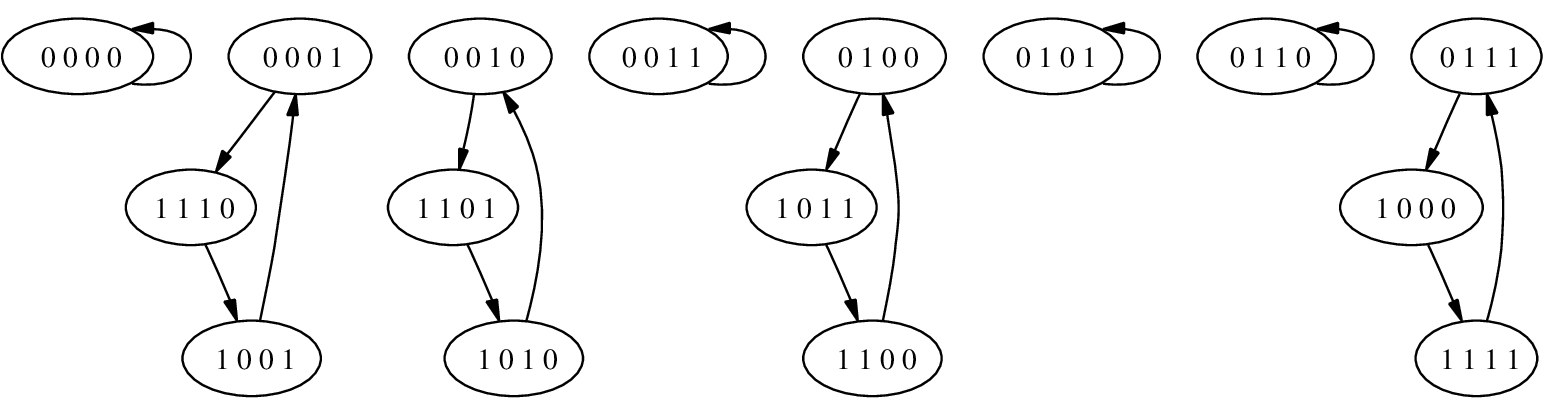}}}}
 \caption{The phase spaces from Example \ref{run-ex-sequential}: $\Phi_\pi$ (left) and  $\Phi_{id}$ (right)}
\label{run-ex-seq}
\end{figure}

\end{example}

Notice that the phase space of any function is a
directed graph in which every vertex has out-degree one; this is
a characteristic property of deterministic functions.

Making use of Boolean arithmetic is a powerful tool in studying
Boolean networks, which is not available in general. In order to
have available an enhanced set of tools it is often natural to
make an additional assumption regarding $X$, namely that it is a
finite number system, a \emph{finite field} \cite{LN}. This
amounts to the assumption that there are ``addition" and
``multiplication" operations defined on $X$ that satisfy the same
rules as ordinary addition and multiplication of numbers. Examples
include $\mathbf Z_p$, the integers modulo a prime $p$. This
assumption can be thought of as the discrete analog of imposing a
coordinate system on an affine space.

When the set $X$ is a finite field, it is easy to show that for
any local function $g$, there exists a polynomial $h$ such that
$g(\vname_1,\dots,\vname_n) = h(\vname_1,\dots, \vname_n)$ for all $(\vname_1,\dots,\vname_n)
\in X^n$. To be precise, suppose $X$ is a finite field with $q$
elements. Then 
\begin{equation}\label{poly-cord}
g(\vname_1,\dots,\vname_n) = \sum_{(c_1,\dots,c_n) \in X^n}g(c_1,\dots,c_n)
\prod_{i=1}^n (1-(\vname_i-c_i)^{q-1}).
\end{equation}
This observation has many useful consequences, since polynomial
functions have been studied extensively and many analytical tools
are available.

Notice that cellular automata and Boolean networks, both parallel
and sequential, are special classes of polynomial dynamical
systems. In fact, it is straightforward to see that
\begin{equation}\label{bool-ploy}
x \wedge y = x\cdot y, \, x \vee y = x+y+xy \mbox{ and } \neg x =
x+1.
\end{equation}
Therefore, the modeling framework of finite dynamical systems
includes that of cellular automata, discussed earlier.  Also,
since a Hopfield network is a function $X^n\longrightarrow X^n$,
which can be represented through its local constituent functions,
it follows that Hopfield networks also are special cases of finite
dynamical systems.

\subsection{Stochastic Finite Dynamical Systems}
The deterministic framework of finite dynamical systems can be
made stochastic in several different ways, making one or more of
the system's defining data stochastic. For example, one could use
one or both of the following criteria.
\begin{itemize}
\item Assume that each variable has a nonempty set of local
functions assigned to it, together with a probability distribution
on this set, and each time a variable is updated, one of these
local functions is chosen at random to update its state. We call
such systems \emph{probabilistic finite dynamical systems} (PFDS),
a generalization of probabilistic Boolean networks
\cite{Shmulevich:02}. \item Fix a subset of permutations $T
\subseteq S_n$ together with a probability distribution. When it
is time for the system to update its state, a permutation $\pi \in
T$ is chosen at random and the agents are updated sequentially
using $\pi$. We call such systems \emph{stochastic finite
dynamical systems} (SFDS).
\end{itemize}

\begin{remark}
By Remark \ref{seq-para}, each system $\Phi_\pi$ is a parallel
system. Hence a SFDS is nothing but a set of parallel dynamical
systems $\{\Phi_\pi \, : \, \pi \in T\}$, together with a
probability distribution. When it is time for the system to update
its state, a system $\Phi_\pi$ is chosen at random and used for
the next iteration.
\end{remark}

To describe the phase space of a stochastic finite dynamical
system, a general method is as follows. Let $\Omega$ be a finite
collection of systems $\Phi_1,\ldots ,\Phi_t$, where $\Phi_i :X^n
\longrightarrow X^n$ for all $i$, and consider the probabilities
$p_1,\ldots ,p_t$ which sum to 1. We obtain the stochastic phase
space
\begin{equation}
\Gamma_\Omega = p_1 \Gamma_1 +  p_2 \Gamma_2 + \cdots  + p_t \Gamma_t  \;,
\end{equation}
where $\Gamma_i$ is the phase space of $\Phi_i$. The associated
probability space is $\mathcal{F} = (\Omega, 2^\Omega,\mu)$, where
the probability measure $\mu$ is induced by the probabilities
$p_i$. It is clear that the stochastic phase space can be viewed
as a \emph{Markov chain} over the state space $X^n$. The adjacency
matrix of $\Gamma_\Omega$ directly encodes the Markov transition
matrix. This is of course not new, and has been done in,
e.g.,\cite{Dawson:74,Vasershtein:69,Shmulevich:02}. But we
emphasize the point that even though SFDS give rise to Markov
chains \emph{our study of SFDS is greatly facilitated by the rich
additional structure} available in these systems.  To understand
the effect of structural components such as the topology of the
dependency graph or the stochastic nature of the update, it is
important to study them not as Markov chains but as SFDS.

\begin{example}
\label{ex:stphasespace} Consider $\Phi_\pi$ and $\Phi_\gamma$ from
Example \ref{run-ex-sequential} and let $\Gamma_\pi$ and
$\Gamma_\gamma$ be their phases spaces as shown in Figure
\ref{run-ex-seq}. Let $p_1 = p_2 = \dfrac{1}{2}$. The phase space
$\dfrac{1}{2}\Gamma_\pi+\dfrac{1}{2}\Gamma_\gamma$ of the stochastic
sequential dynamical system obtained from $\Phi_\pi$ and
$\Phi_\gamma$ (with equal probabilities) is presented in Figure
\ref{fig:stphasespace}.
\begin{figure}[ht]
\centerline{
\framebox{\includegraphics[width=0.5\textwidth]{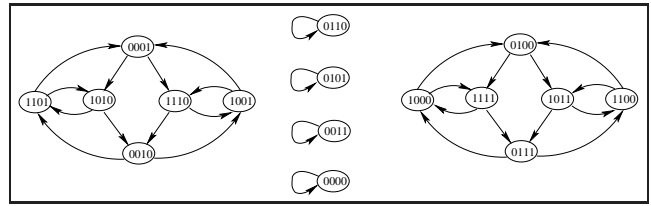}} }
\caption{The stochastic phase space for
Example~\ref{ex:stphasespace} induced by the two deterministic
phase spaces of $\Phi_\pi$ and $\Phi_\gamma$ from Figure
\ref{run-ex-seq}. For simplicity the weights of the edges have
been omitted.} \label{fig:stphasespace}
\end{figure}
\end{example}

\subsection{Agent-based Simulations as Finite Dynamical Systems}

In the following we describe the generic structure of the systems
typically modeled and studied through agent-based simulations. The
central notion is naturally that of an \emph{agent}.

Each agent carries a state that may encode its preferences,
internal configuration, perception of its environment, and so on.
In the case of \transims, for instance, the agents
are the cells making up the road network. The cell state
contains information about whether or not the cell is occupied by a
vehicle as well as the velocity of the vehicle.  One may
assume that each cell takes on states from the same set of
possible states, which may be chosen to support the structure of a
finite field.

The agents interact with each other, but typically an agent only
interacts with a small subset of agents, its \emph{neighbors}.
Through such an interaction an agent may decide to change its
state based on the states (or other aspects) of the agents with
which it interacts. We will refer to the process where an agent
modifies its state through interaction as an \emph{agent update}.
The precise way in which an agent modifies its state is governed
by the nature of the particular agent.  In \transims the neighbors
of a cell are the adjacent road network cells.  From this
adjacency relation one obtains a dependency graph of the agents.
The local update function for a given agent can be obtained from
the rules governing traffic flow between cells.

The updates of all the agents may be scheduled in different ways.
Classical approaches include synchronous, asynchronous and
event-driven schemes. The choice will depend on system properties
or particular considerations about the simulation implementation.

In the case of \emph{CImmSim}, the situation is somewhat more
complicated.  Here the agents are also the spatial units of the
system, each representing a small volume of lymph tissue.
 The total volume is represented as a 2-dimensional CA, in which
every agent has 4 neighbors, so that the dependency graph is a
regular 2-dimensional grid. The state of each agent is a
collection of counts for the various immune cells and pathogens
that are present in this particular agent (volume). Movement
between cells is implemented as diffusion. Immune cells can
interact with each other and with pathogens while they reside in
the same volume. Thus, the local update function for a given cell
of the simulation is made up of the two components of movement
between cells and interactions within a cell.  For instance, a B
cell could interact with the Epstein-Barr virus in a given volume
and transition from uninfected to infected by the next time step.
Interactions as well as movement are stochastic, resulting in a
stochastic finite dynamical system.  The update order is parallel.

\begin{example}[The voting game (Section~\ref{sec:voting})] The
following scenario is constructed to illustrate how implementation
choices for the system components have a clear and direct bearing on the
dynamics and simulation outcomes.

Let the voter dependency graph be the star graph on $5$ vertices with
center vertex $a$ and surrounding vertices $b$, $c$, $d$ and
$e$. Furthermore, assume that everybody votes opportunistically
using the majority rule: the vote cast by an individual is equal to
the preference of the majority of his/her friends with the person's own
preference included. For simplicity, assume candidate~$1$ is preferred
in the case of a tie.

If the initial preference is $x_a = 1$ and $x_b = x_c = x_d = x_e
= 0$ then if voter $a$ goes first he/she will vote for
candidate~$0$ since that is the choice of the majority of the
neighbor voters. However, if $b$ and $c$ go first they only know
$a$'s preference. Voter $b$ therefore casts his/her vote for
candidate $1$ as does $c$. Note that this is a tie situation with
an equal number of preferences for candidate $1$ ($a$) and for
candidate $2$ ($b$). If voter $a$ goes next, then the situation
has changed: the preference of $b$ and $c$ has already changed to
$1$. Consequently, voter $a$ picks candidate $1$. At the end of
the day candidate~$1$ is the election winner, and the choice of
update order has tipped the election!

This example is of course constructed to illustrate our point.
However, in real systems it can be much more difficult to detect
the presence of such sensitivities and their implications. A solid
mathematical framework can be very helpful in detecting such
effects.

\end{example}

\section{Finite dynamical systems as theoretical and computational tools}

If finite dynamical systems are to be useful as a modeling
paradigm for agent-based simulations it is necessary that they can
serve as a fairly universal model of computation. We discuss here
how such dynamical systems can mimic Turing Machines (TMs), a
standard universal model for computation. For a more thorough
exposition, we refer the reader to the series of papers by Barrett
et al. \cite{BH+dichotomy,BH+pred-tcs,BH+Annals,BH+TCS,BH+ijcai}.
To make the discussion reasonably self-contained, we provide a brief and
informal discussion of the TM model. Additional information on TMs
can be found in any standard text on the theory of computation
(e.g. \cite{Sip97}).

\subsection{A Computational View of Finite Dynamical Systems:
Definitions}

In order to understand the relationship of finite dynamical
systems to TMs, it is important to view such systems from a
computational stand point. Recall that a finite dynamical system
$\Phi:X^n\longrightarrow X^n$, where $X$ is a finite set, has an underlying
dependency graph $Y(V,E)$. From a computational point of view, the
nodes of the dependency graph (the agents in the simulation) are
thought of as devices that compute appropriate functions. For
simplicity, we will assume in this section that the dependency
graph is undirected, that is, all dependency relations are
symmetric. At any time, the state value of each node $v_i \in V$
is from the specified domain $X$. The inputs to
$f_i$ are the current state of $v_i$ and the states of the
neighbors of $v_i$ as specified by $Y$. The output of $f_i$, which
is also a member of $X$, becomes the state of $v_i$ at the next
time instant. The discussion in this section will focus on
sequentially updated systems (SDS), but all results discussed
apply to parallel systems as well.

Each step of the computation carried out by an SDS can be thought
as consisting of $n$ ``mini steps;" in each mini~step, the value
of the local transition function at a node is computed and the
state of the node is changed to the computed value. Given an SDS
$\Phi$, a \emph{configuration} $\mathcal C$ of $\Phi$ is a
vector $(c_1, c_2, \ldots, c_n)\in X^n$. It can be seen that each
computational step of an SDS causes a transition from one
configuration to another.

\subsection{Configuration Reachability Problem for SDSs}

Based on the computational view, a number of different problems
can be defined for SDSs (see for example,
\cite{BH+dichotomy,BH+pred-tcs,BH+ge_fix}). To illustrate how SDSs
can model TM computations, we will focus on one such problem,
namely the \emph{configuration reachability} (CR) problem: Given
an SDS $\Phi$, an initial configuration $\mathcal C$ and another
configuration $\mathcal C'$, will $\Phi$, starting from $\mathcal
C$, ever reach configuration $\mathcal C'$? The problem can also
be expressed in terms of the phase space of $\Phi$. Since
configurations such as $\mathcal C$ and $\mathcal C'$ are
represented by nodes in the phase space, the CR problem boils down
to the question of whether there is a directed path in the phase
space from $\mathcal C$ to $\mathcal C'$. This abstract
problem can be mapped to several problems in the simulation of
multiagent systems. Consider for example the \transims context.
Here, the initial configuration $\mathcal C$ may represent the
state of the system early in the day (when the traffic is very
light) and $\mathcal C'$ may represent an ``undesirable" state of
the system (such as heavy traffic congestion). Similarly, in the
context of modeling an infectious disease, $\mathcal C$ may
represent the initial onset of the disease (when only a small
number of people are infected) and $\mathcal C'$ may represent a
situation where a large percentage of the population is infected.
The purpose of studying computational problems such as CR is to
determine whether one can efficiently predict the occurrence of
certain events in the system from a description of the system. If
computational analysis shows that the system can indeed reach
undesirable configurations as it evolves, then one can try to
identify steps needed to deal with such situations.

\subsection{Turing Machines: A Brief Overview}

A Turing machine (TM) is a simple and commonly used model for
general purpose computational devices. Since our goal is to point
out how SDSs can also serve as computational devices, we will
present an informal overview of the TM model. Readers interested
in a more formal description may consult~\cite{Sip97}.

A TM consists of a set $Q$ of states, a one-way infinite input
tape and a read/write head that can read and modify symbols on the
input tape. The input tape is divided into cells and each cell
contains a symbol from a special finite alphabet. An input
consisting of $n$ symbols is written on the leftmost $n$ cells of
the input tape.  (The other cells are assumed to contain a special
symbol called \emph{blank}.) One of the states in $Q$, denoted
by $q_s$, is the designated \emph{start} state. $Q$ also
includes two other special states, denoted by $q_a$ (the
\emph{accepting} state) and $q_r$ (the \emph{rejecting}
state). At any time, the machine is in one of the states in $Q$.
The \emph{transition function} for the TM specifies for each
combination of the current state and the current symbol under the
head, a new state, a new symbol for the current cell (which is
under the head) and a movement (i.e., left or right by one cell)
for the head. The machine starts in state $q_s$ with the head on
the first cell of the input tape. Each step of the machine is
carried out in accordance with the transition function. If the
machine ever reaches either the accepting or the rejecting state,
it halts with the corresponding decision; otherwise, the machine
runs forever.

A \emph{configuration} of a TM consists of its current state,
the current tape contents and the position of the head. Note that
the transition function of a TM specifies how a new configuration
is obtained from the current configuration.

The above description is for the basic TM model (also called
\emph{single tape} TM model). For convenience in describing some
computations, several variants of the above basic model have been
proposed. For example, in a \emph{multi-tape} TM, there are one
or more \emph{work tapes} in addition to the input tape. The
work tapes can be used to store intermediate results. Each work
tape has its own read/write head and the definitions of
configuration and transition function can be suitably modified to
accommodate work tapes. While such an enhancement to the basic TM
model makes it easier to carry out certain computations, it does
not add to the machine's computational power. In other words, any
computation that can be carried out using the enhanced model can
also be carried out using the basic model.

As in the case of dynamical systems, one can define a
\emph{configuration reachability} (CR) problem for TMs: Given a TM
$M$, an initial configuration $\cali_M$ and another configuration
$\calc_M$, will the TM starting from $\cali_M$ ever reach $\calc_M$?
We refer to the CR problem in the context of TMs as CR-TM. In fact, it
is this problem for TMs that captures the essence of what can be
effectively computed. In particular, by choosing the state component
of $\calc_M$ to be one of the halting states ($q_a$ or $q_r$), the
problem of determining whether a function is computable is transformed
into an appropriate CR-TM problem. By imposing appropriate
restrictions on the resources used by a TM (e.g. the number of steps,
the number of cells on the work tapes), one obtains different versions
of the CR-TM problem which characterize different computational
complexity classes~\cite{Sip97}.

\subsection{How SDSs Can Mimic Turing Machines}

The above discussion points out an important similarity between
SDSs and TMs. Under both of these models, each computational step
causes a transition from one configuration to another. It is this
similarity that allows one to construct a discrete dynamical
system $\Phi$ that can simulate a TM. Typically, each step of a TM
is simulated by a short sequence of successive iterations $\Phi$.
As part of the construction, one also identifies a suitable
mapping between the configurations of the TM being simulated and
those of the dynamical system. This is done in such a way that the
answer to CR-TM problem is ``yes'' if and only if the answer to
the CR problem for the dynamical system is also ``yes.''

To illustrate the basic ideas, we will informally sketch a
construction from \cite{BH+dichotomy}. This construction produces
an SDS $\Phi$ that simulates a restricted version of TMs; the
restriction being that for any input containing $n$ symbols, the
number of work tape cells that the machine may use is bounded by a
linear function of $n$. Such a TM is called a \emph{linear
bounded automaton} (LBA) \cite{Sip97}. Let $M$ denote the given
LBA and let $n$ denote the length of the input to $M$. The domain
$X$ for the SDS $\Phi$ is chosen to be a finite set based on the
allowed symbols in the input to the TM. The dependency graph is
chosen to be a simple path on $n$ nodes, where each node serves as
a representative for a cell on the input tape. The initial and
final configurations $\mathcal C$ and $\mathcal C'$ for $\Phi$ are
constructed from the corresponding configurations of $M$. The
local transition function for each node of the SDS is constructed
from the given transition function for $M$ in such a way that each
step of $M$ corresponds to exactly one step of $\Phi$. Thus, there
is a simple bijection between the sequence of configurations that
$M$ goes through during its computation and the sequence of states
that $\Phi$ goes through as it evolves. Using this bijection, it
is shown in \cite{BH+dichotomy} that the answer to the CR-TM
problem is ``yes" if and only if $\Phi$ reaches $\mathcal C'$
starting from $\mathcal C$. Reference \cite{BH+dichotomy} also
presents a number of sophisticated constructions where the
resulting dynamical system is very simple; for example, it is
shown that an LBA can be simulated by an SDS in which $X$ is the
Boolean field, the dependency graph is $d$-regular for some
constant $d$ and the local transition functions at all the nodes
are \emph{identical}.  Such results point out that one does not
need complicated dynamical systems to model TM computations.

Barrett et al. \cite{BH+ijcai} have also considered
\emph{stochastic} SDS (SSDS), where the local transition
function at each node is stochastic. For each combination
of inputs, a stochastic local transition function specifies a
probability distribution over the domain of state values. It is
shown in \cite{BH+ijcai} that SSDSs can effectively simulate
computations carried out by probabilistic TMs (i.e., TMs whose
transition functions are stochastic).

\subsection{TRANSIMS related questions}
Section~\ref{sec:transims} gave an overview of some aspects of the
\transims model. The micro-simulator module is specified as a
functional composition of four cellular automata of the form
$\Delta_4 \circ \Delta_3 \circ \Delta_2 \circ \Delta_1 $. (We only
described $\Delta_3$ which corresponds to velocity updates.) Such
a formulation has several advantages. First, it has created an
abstraction of the essence of the system in a precise mathematical
way without burying it in contextual domain details. An immediate
advantage of this specification is that it makes the whole
implementation process more straightforward and transparent. While
the local update functions for the cells are typically quite
simple, for any realistic study of an urban area the problem size
would typically require a sequential implementation, raising a
number of issues that are best addressed within a mathematical
framework like the one considered here.

\section{Mathematical results on finite dynamical systems}

In this section we outline a collection of mathematical results
about finite dynamical systems that is representative of the
available knowledge.  The majority of these results are about
deterministic systems, as the theory of stochastic systems of this
type is still in its infancy.
We will first consider synchronously updated systems.

Throughout this section, we make the assumption that the state set
$X$ carries the algebraic structure of a finite field.
Accordingly, we use the notation $\kk$ instead of $X$.  It is a
standard result that in this case the number $q$ of elements in
$\kk$ has the form $q = p^t$ for some prime $p$ and $t \geq 1$.
The reader may keep the simplest case $\kk = \{0, 1\}$ in mind, in
which case we are considering Boolean networks.

Recall Equation (\ref{poly-cord}). That is,
any function $g: \kk^n\longrightarrow \kk$ can be
represented by a multivariate polynomial with coefficients in
$\kk$.  If we require that the exponent of each variable be less
than $q$, then this representation is unique.
In particular Equation (\ref{bool-ploy})
implies that every Boolean function can be represented uniquely as a
polynomial function.

\subsection{Parallel update systems} Certain classes of finite
dynamical systems have been studied extensively, in particular
cellular automata and Boolean networks.  Many of these studies
have been experimental in nature, however.  Some more general
mathematical results about cellular automata can be found in the
papers of Wolfram and collaborators \cite{Wolfram:86a}.  The
results there focus primarily on one-dimensional Boolean cellular
automata with a particular fixed initial state.  Here we collect a
sampling of more general results.

We first consider linear and affine systems
\[
\Phi = (f_1, \ldots ,f_n): \kk^n\longrightarrow \kk^n
\]
That is, we consider systems for which the coordinate
functions $f_i$ are linear, resp. affine, polynomials. (In the
Boolean case this includes functions constructed from XOR and
negation.) When each $f_i$ is a linear polynomial of the form
$f_i(x_1,\dots,x_n) = a_{i1}x_1+ \cdots + a_{in}x_n$, the map $\Phi$
is nothing but a \emph{linear transformation} on $\kk^n$ over
$\kk$, and, by using the standard basis, $\Phi$ has the matrix
representation
\[
\Phi \left(\left[%
\begin{array}{c}
  x_1 \\
  \vdots \\
  x_n \\
\end{array}%
\right]\right) = \left[
\begin{array}{ccc}
  a_{11} & \cdots & a_{1n} \\
  \vdots & \ddots & \vdots \\
  a_{n1} & \cdots & a_{nn} \\
\end{array}
\right] \left[
\begin{array}{c}
  x_1 \\
  \vdots \\
  x_n \\
\end{array}%
\right],
\]
where $a_{ij} \in \kk$ for all $i,j$.

Linear finite dynamical systems were first studied by
Elspas~\cite{El}. His motivation came from studying feedback shift
register networks and their applications to radar and
communication systems and automatic error correction circuits. For
linear systems over finite fields of prime cardinality, that is,
$q$ is a prime number, Elspas showed that the exact number and
length of each limit cycle can be determined from the
\textit{elementary divisors} of the matrix $A$. Recently,
Hernandez \citep{He} re-discovered Elspas' results and generalized
them to arbitrary finite fields. Furthermore, he showed that the
structure of the \textit{tree of transients} at each node of each
limit cycle is the same, and can be completely determined from the
nilpotent elementary divisors of the form $x^a$. For
\textit{affine} Boolean networks (that is, finite dynamical
systems over the Boolean field with two elements, whose local
functions are linear polynomials which might have constant terms),
a method to analyze their cycle length has been developed in
\citep{MW}. After embedding the matrix of the transition function,
which is of dimension $n \times (n+1)$, into a square matrix of
dimension $n+1$, the problem is then reduced to the linear case. A
fast algorithm based on \cite{He} has been implemented in
\cite{JLSV}, using the symbolic computation package
\emph{Macaulay2}.

It is not surprising that the phase space structure of $\Phi$ should
depend on invariants of the matrix $A = (a_{ij})$.  The rational
canonical form of $A$ is a block-diagonal matrix and one can
recover the structure of the phase space of $A$ from that of the
blocks in the rational form of $A$.  Each block represents either an
invertible or a nilpotent linear transformation.  Consider an
invertible block $B$.  If  $\mu(x)$ is the
minimal polynomial of $B$, then there exists $s$ such that
$\mu(x)$ divides $x^s-1$. Hence $B^s - I = 0$ which implies that $B^s {\bf
v}= {\bf v}$. That is, every state vector ${\bf v}$ in the phase
space of $B$  is in a cycle whose length is a divisor of $s$.

\begin{definition}
For any polynomial $\lambda(x)$ in $\kk[x]$, the \emph{order} of
$\lambda(x)$ is the least integer $s$ such that $\lambda(x)$
divides $x^s-1$.
\end{definition}

The cycle structure of the phase space of $\Phi$ can be completely
determined from the orders of the irreducible factors of the minimal
polynomial of $\Phi$.  The computation of these orders involves in
particular the factorization of numbers of the form $q^r-1$, which
makes the computation of the order of a polynomial potentially quite
costly.  The nilpotent blocks in the decomposition of $A$ determine
the tree structure at the nodes of the limit cycles.  It turns out
that all trees at all periodic nodes are identical.  This generalizes
a result in~\cite{Martin:84} for additive cellular automata over the
field with two elements.

While the fact that the structure of the phase space of a linear
system can be determined from the invariants associated with its
matrix may not be unexpected, it is a beautiful example of how the
right mathematical viewpoint provides powerful tools to completely
solve the problem of relating the structure of the local functions
to the resulting (or emerging) dynamics.  Linear and affine
systems have been studied extensively in several different
contexts and from several different points of view, in particular
the case of cellular automata. For instance, additive cellular
automata over more general rings as state sets have been studied,
e.g., in \cite{Chaudhuri:97a}. Further results on additive CAs can
also be found there. One important focus in \cite{Chaudhuri:97a}
is on the problem of finding CAs with limit cycles of maximal
length for the purpose of constructing pseudo random number
generators.

Unfortunately, the situation is more complicated for nonlinear
systems. For the special class of Boolean synchronous systems whose
local update functions consist of monomials, there is a polynomial time
algorithm that determines whether a given monomial system has only
fixed points as periodic points~\cite{CLP}. This question was
motivated by applications to the modeling of biochemical networks.
The criterion is given in terms of invariants of the dependency graph
$Y$.  For a strongly connected directed graph $Y$ (i.e., there is a
directed path between any pair of vertices), its {\it loop number} is the
greatest common divisor of all directed loops at a particular
vertex. (This number is independent of the vertex chosen.)

\begin{theorem}[\cite{CLP}]
A Boolean monomial system has only fixed points as periodic points if
and only if the loop number of every strongly connected component of
its dependency graph is equal to 1.
\end{theorem}

In \cite{CJLS} it is shown that the problem for general finite fields
can be reduced to that of a Boolean system and a linear system over
rings of the form $\zz/p^r\zz$, $p$ prime. Boolean monomial systems
have been studied before in the cellular automaton context~\cite{BG}.

\subsection{Sequential update systems}
The update order in a sequential dynamical system has been studied
using combinatorial and algebraic techniques. A natural question to
ask here is how the system depends on the update schedule.
In~\cite{Mortveit:01a,Barrett:01a,Barrett:00a,Reidys:98a} this was
answered on several levels for the special case where the update
schedule is a permutation. We describe these results in some detail.
Results about the more general case of update orders described by
words on the indices of the local update functions can be found
in~\cite{GJL}.

Given local update functions $f_i: k^n \longrightarrow k$ and
permutation update orders $\sigma, \pi$, a natural question is
when the two resulting SDS $\Phi_\sigma$ and $\Phi_\pi$ are
identical and, more generally, how many different systems one
obtains by varying the update order over all permutations.  Both
questions can be answered in great generality. The answer involves
invariants of two graphs, namely the acyclic orientations of the
dependency graph $Y$ of the local update functions and the
\emph{update graph} of $Y$. The update graph $U(Y)$ of $Y$ is the
graph whose vertex set consists of all permutations of the vertex
set of $Y$~\cite{Reidys:98a}. There is an (undirected) edge
between two permutations $\sigma = (\sigma_1,\ldots ,\sigma_n)$
and $\tau = (\tau_1, \ldots ,\tau_n)$
 if they differ by a transposition of two
adjacent entries $\sigma_i$ and $\sigma_{i+1}$ such that there is
no edge in $Y$ between $\sigma_i$ and $\sigma_{i+1}$.

The update graph encodes the fact that one can commute two local
update functions $f_i$ and $f_j$ without affecting the end result
$\Phi$ if $i$ and $j$ are not connected by an edge in $Y$.  That
is, $\cdots f_i\circ f_j\cdots = \cdots f_j\circ f_i\cdots$ if and
only if $i$ and $j$ are not connected by an edge in $Y$.

All permutations belonging to the same connected component in
$U(Y)$ give identical \sds{} maps. The number of (connected)
components in $U(Y)$ is therefore an upper bound for the number of
functionally inequivalent \sds{} that can be generated by just
changing the update order. This can also be characterized in terms
of acyclic orientations of the graph $Y$: each component in the
update graph induces a unique acyclic orientation of the graph
$Y$. Moreover, we have the following result: 
\begin{proposition}[\cite{Reidys:98a}]
There is a bijection
\begin{equation*}
F_Y \colon S_Y /\sim_Y \longrightarrow \operatorname{Acyc}(Y) \;,
\end{equation*}
where $S_Y/\sim_Y$ denotes the set of connected components of
$U(Y)$ and $\operatorname{Acyc}(Y)$ denotes the set of acyclic
orientations of $Y$.
\end{proposition}

This upper bound on the number of functionally different systems
has been shown in \cite{Reidys:98a} to be sharp for Boolean
systems, in the sense that for a given $Y$ one construct this
number of different systems, using appropriate combinations of NOR
functions.

For two permutations $\sigma$ and $\tau$ it is easy to determine if
they give identical \sds{} maps: one can just compare their induced
acyclic orientations. The number of acyclic orientations of the graph
$Y$ tells how many functionally different \sds{} maps one can obtain for
a fixed graph and fixed vertex functions.  The work of Cartier and
Foata~\cite{Cartier:69} on partially commutative monoids studies a
similar question, but their work is not concerned with finite
dynamical systems.

Note that permutation update orders have been studied sporadically
in the context of cellular automata on circle
graphs~\cite{Park:86} but not in a systematic way, typically using
the order $(1,2,\dots,n)$ or the even-odd/odd-even orders. As a
side note, we remark that this work also confirms our findings
that switching from a parallel update order to a sequential order
turns the complex behavior found in Wolfram's ``class III and IV''
automata into much more regular or mundane dynamics, see
e.g.~\cite{Schoenfisch:99}.

The work on functional equivalence was extended to dynamical
equivalence (topological conjugation)
in~\cite{Mortveit:01a,Barrett:01a}.  The automorphism group of the
graph $Y$ can be made to act on the components of the update graph
$U(Y)$ and therefore also on the acyclic orientations of $Y$. All
permutations contained in components of an orbit under
$\operatorname{Aut}(Y)$ give rise to dynamically equivalent
sequential dynamical systems, that is, to isomorphic phase spaces.
 However, here one needs some more technical assumptions, i.e., the
local functions must be symmetric and induced, see~\cite{BMR3}.
This of course also leads to a bound for the number of dynamically
inequivalent systems that can be generated by varying the update
order alone. Again, this was first done for permutation update
orders. The theory was extended to words over the vertex set of
$Y$ in \cite{Reidys:04a,GJL}.

The structure of the graph $Y$ influences the dynamics of the
system.  As an example, graph invariants such as the independent
sets of $Y$ turn out to be in a bijective correspondence with the
invariant set of sequential systems over the Boolean field $\kk$
that have $\operatorname{nor}_t\colon \kk^t \to \kk_2$ given by
$\operatorname{nor}_t(x_1, \dots, x_t) = (1+x_1)\cdots(1+x_t)$ as
local functions~\cite{Reidys:01a}. This can be extended to other
classes such as those with order independent invariant sets as
in~\cite{Hansson:05b}. We have already seen how the automorphisms
of a graph give rise to equivalence~\cite{Mortveit:01a}. Also, if
the graph $Y$ has non-trivial covering maps we can derive
simplified or reduced (in an appropriate sense) versions of the
original SDS over the image graphs of $Y$, see
e.g.~\cite{Mortveit:04b,Reidys:03a}.

Parallel and sequential dynamical systems differ when it comes to
invertibility. Whereas it is generally computationally intractable
to determine if a CA over $\zz^d$ is invertible for
$d\ge2$~\cite{Kari:05}, it is straightforward to determine this
for a sequential dynamical system~\cite{Mortveit:01a}. For
example, it turns out that the only invertible Boolean sequential
dynamical systems with symmetric local functions are the ones
where the local functions are either the parity function or the
logical complement of the parity function~\cite{Barrett:01a}.

Some classes of sequential dynamical systems such as the ones
induced by the $\operatorname{nor}$-function have desirable
stability properties~\cite{Hansson:05b}. These systems have
minimal invariant sets (i.e. periodic states) that do not depend
on the update order. Additionally, these invariant sets are stable
with respect to configuration perturbations.

If a state $\mathbf c$ is perturbed to a state $\mathbf c'$ that
is not periodic this state will evolve to a periodic state
$\mathbf c''$ in one step; that is, the system will quickly return
to the invariant set. However, the states $\mathbf c$ and $\mathbf
c''$ may not necessarily be in the same periodic orbit.


\subsection{The category of sequential dynamical systems}

As a general principle, in order to study a given class of
mathematical objects it is useful to study transformations between
them.  In order to provide a good basis for a mathematical analysis
the objects and transformations together should form a {\it category},
that is, the class of transformations between two objects should
satisfy certain reasonable properties (see, e.g.,~\cite{MacLane:98}).
Several proposed definitions of a transformation of SDS have been
published, notably in~\cite{LP2} and~\cite{Reidys:03a}.  One possible
interpretation of a transformation of SDS from the point of view of
agent-based simulation is that the transformation represents the
approximation of one simulation by another or the embedding/projection
of one simulation into/onto another.  These concepts have obvious
utility when considering different simulations of the same complex
system.

One can take different points of view in defining a transformation
of SDS.  One approach is to require that a transformation is
compatible with the defining structural elements of an SDS, that
is, with the dependency graph, the local update functions, and the
update schedule.  If this is done properly, then one should expect
to be able to prove that the resulting transformation induces a
transformation at the level of phase spaces.  That is,
transformations between SDS should preserve the local and
global dynamic behavior.  This implies that transformations
between SDS lead to transformations between the associated global
update functions.

Since the point of view of SDS is that global dynamics emerges
from system properties that are defined locally, the notion of SDS
transformation should focus on the local structure.  This is the
point of view taken in~\cite{LP2}. The definition given there is
rather technical and the details are beyond the scope of this
article.  The basic idea is as follows.  Let $\Phi_\pi =
f_{\pi(n)} \circ \cdots \circ f_{\pi(1)}$ and $\Phi_\sigma =
g_{\sigma(m)} \circ \cdots \circ g_{\sigma(1)}$ with the
dependency graphs $Y_\pi$ and $Y_\gamma$, respectively. A
transformation $F: \Phi_\pi \longrightarrow \Phi_\sigma$ is
determined by 
\begin{itemize}
\item a graph mapping $\varphi :Y_\pi\longrightarrow Y_\gamma $
(reverse direction); \item a family of maps
$k_{\phi(v)}\longrightarrow k_v, v\in Y_\pi$;  \item an order
preserving map $\sigma \longmapsto \pi$ of update schedules.
\end{itemize}
These maps are required to satisfy the property that they ``locally''
assemble to a coherent transformation.
Using this definition of transformation, it is shown~\cite[Theorem
2.6]{LP2} that the class of SDS
forms a category. One of the requirements, for instance, is that the
composition of two transformations is again a
transformation. Furthermore, it is shown~\cite[Theorem 3.2]{LP2} that
a transformation of SDS induces a map of directed graphs on the phase
spaces of the two systems.  That is, a transformation of the local
structural elements of SDS induces a transformation of global
dynamics.  One of the results proven in~\cite{LP2} is that every SDS
can be decomposed uniquely into a direct product (in the categorical
sense) of indecomposable SDS.

Another possible point of view is that a transformation
\[
F: (\Phi_\pi:k^n\rightarrow k^n)\longrightarrow
(\Phi_\gamma:k^m\rightarrow k^m)
\]
is a function $F: k^n\longrightarrow k^m$ such that $F \circ
\Phi_\pi = \Phi_\gamma \circ F$, without requiring specific
structural properties.  This is the approach in~\cite{Reidys:03a}.
This definition also results in a category, and a collection of
mathematical results. Whatever definition chosen, much work
remains to be done in studying these categories and their
properties.

\section{Future directions}
Agent-based computer simulation is an important method for
modeling many complex systems, whose global dynamics emerges from
the interaction of many local entities.  Sometimes this is the
only feasible approach, especially when available information is not enough
to construct global dynamic models.  The size of many realistic
systems, however, leads to computer models that are themselves
highly complex, even if they are constructed from simple software
entities.  As a result it becomes challenging to carry out
verification, validation, and analysis of the models, since these
consist in essence of complex computer programs.  This chapter
argues that the appropriate approach is to provide a formal
mathematical foundation by introducing a class of mathematical
objects to which one can map agent-based simulations.  These
objects should capture the key features of an agent-based
simulation and should be mathematically rich enough to allow the
derivation of general results and techniques.  The mathematical
setting of dynamical systems is a natural choice for this purpose.

The class of finite dynamical systems over a state set $X$ which
carries the structure of a finite field satisfies all these
criteria.  Parallel, sequential, and stochastic versions of these
are rich enough to serve as the mathematical basis for models of a
broad range of complex systems. While finite dynamical systems
have been studied extensively from an experimental point of view,
their mathematical theory should be considered to be in its
infancy, providing a fruitful area of research at the interface of
mathematics, computer science, and complex systems theory.


\section{Bibliography}

\bibliographystyle{siam}
\renewcommand{\refname}{Primary Litreture}
\bibliography{math_agent_based_models}
\renewcommand{\refname}{Books and Reviews}

\end{document}